\PassOptionsToPackage{bookmarks=false}{hyperref}
\documentclass[sigconf, nonacm]{acmart}
\usepackage{flushend}
\usepackage{balance}
\usepackage{mathtools}
\usepackage{dsfont}

\AtBeginDocument{%
  \providecommand\BibTeX{{%
    \normalfont B\kern-0.5em{\scshape i\kern-0.25em b}\kern-0.8em\TeX}}}

\setcopyright{acmcopyright}

\copyrightyear{2020}
\acmYear{2020}
\setcopyright{acmcopyright}
\acmConference[ICSE '20]{42nd International Conference on Software Engineering}{May 23--29, 2020}{Seoul, Republic of Korea}
\acmBooktitle{42nd International Conference on Software Engineering (ICSE '20), May 23--29, 2020, Seoul, Republic of Korea}
\acmPrice{15.00}
\acmDOI{10.1145/3377811.3380339}
\acmISBN{978-1-4503-7121-6/20/05}

\setcopyright{none}
\renewcommand\footnotetextcopyrightpermission[1]{} 

\begin{document}

\title{Structure-Invariant Testing for Machine Translation}

\author{Pinjia He}
\orcid{0000-0003-3377-8129}             
\affiliation{
  \department{Department of Computer Science}              
  \institution{ETH Zurich}            
  \country{Switzerland}                    
}
\email{pinjia.he@inf.ethz.ch}          

\author{Clara Meister}
\affiliation{
  \department{Department of Computer Science}              
  \institution{ETH Zurich}            
  \country{Switzerland}                    
}
\email{clara.meister@inf.ethz.ch }          

\author{Zhendong Su}
\affiliation{
  \department{Department of Computer Science}              
  \institution{ETH Zurich}            
  \country{Switzerland}                    
}
\email{zhendong.su@inf.ethz.ch }          


\begin{abstract}
In recent years, machine translation software has increasingly been integrated into our daily lives. People routinely use machine translation for various applications, such as describing symptoms to a foreign doctor and reading political news in a foreign language. However, the complexity and intractability of neural machine translation (NMT) models that power modern machine translation make the robustness of these systems difficult to even assess, much less guarantee. Machine translation systems can return inferior results that lead to misunderstanding, medical misdiagnoses, threats to personal safety, or political conflicts. Despite its apparent importance, validating the robustness of machine translation systems is very difficult and has, therefore, been much under-explored.

To tackle this challenge, we introduce \textit{structure-invariant testing (SIT)}, a novel metamorphic testing approach for validating machine translation software. Our key insight is that the translation results of ``similar'' source sentences should typically exhibit similar sentence structures. Specifically, SIT (1) generates similar source sentences by substituting one word in a given sentence with semantically similar, syntactically equivalent words; (2) represents sentence structure by syntax parse trees (obtained via constituency or dependency parsing); (3) reports sentence pairs whose structures differ quantitatively by more than some threshold. To evaluate SIT, we use it to test Google Translate and Bing Microsoft Translator with 200 source sentences as input, which led to 64 and 70 buggy issues with 69.5\% and 70\% top-1 accuracy, respectively. The translation errors are diverse, including under-translation, over-translation, incorrect modification, word/phrase mistranslation, and unclear logic. 
\vspace{-1ex}


\end{abstract}



\vspace{-2ex}


\maketitle

\section{Introduction}\label{sec:intro}

Machine translation software has seen rapid growth in the last decade; users now rely on machine translation for a variety of applications, such as signing lease agreements when studying abroad, describing symptoms to a foreign doctor, and reading political news in a foreign language.  In 2016, Google Translate, the most widely-used online translation service, attracted more than 500 million users and translated more than 100 billion words per day~\cite{userdata}. On top of this, machine translation services are also embedded into various software applications, such as Facebook~\cite{tranFacebook} and Twitter~\cite{tranTwitter}.

The advances in machine translation that are responsible for such growth can largely be attributed to neural machine translation (NMT) models, which have become the core component of many machine translation systems. As reported by research from Google \cite{Wu16Arxiv} and Microsoft~\cite{Hassan18Arxiv}, state-of-the-art NMT models are approaching human-level performance in terms of accuracy, i.e., BLEU \cite{Papineni02BLEU}. These recent breakthroughs have led users to start relying on machine translation software (e.g., Google Translate~\cite{googletranslate} and Bing Microsoft Translator~\cite{bingtranslator}) in their daily lives. 

\begin{figure*}[t]
\centering{} 
\includegraphics[scale=0.82]{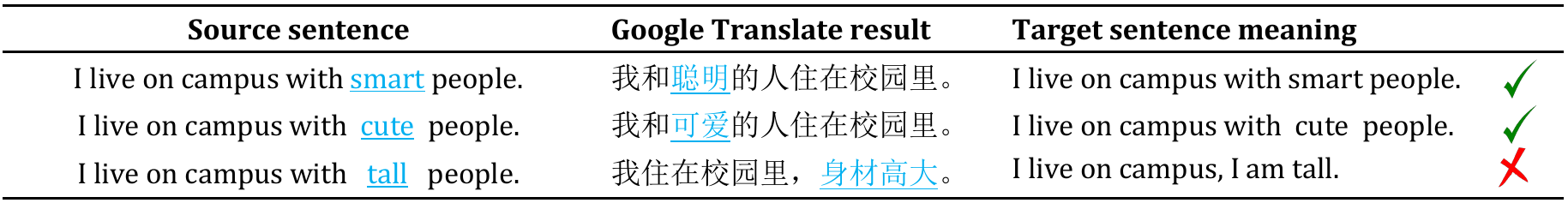}
\caption{Examples of similar source sentences and Google Translate results.}
\label{fig:sitinsight}
\end{figure*}

However, NMT models are not as reliable as many may believe. Recently, sub-optimal and incorrect outputs have been found in various software systems with neural networks as their core components. Typical examples include autonomous cars~\cite{Pei17SOSP, Tian18ICSE, Evtimov18CVPR}, sentiment analysis tools~\cite{Alzantot18EMNLP, Iyyer18NAACL, Li19NDSS}, and speech recognition services~\cite{Carlini16Security,Qin18ICML}. These recent research efforts show that neural networks can easily return inferior results (e.g., wrong class labels) given specially-crafted inputs (i.e., adversarial examples). 
NMT models are no exception; they can be fooled by adversarial examples~\cite{Ebrahimi18COLING} or natural noise (e.g., typos in input sentences)~\cite{Belinkov18ICLR}. These inferior results (i.e., sub-optimal or incorrect translations), can lead to misunderstanding, embarrassment, financial loss, medical misdiagnoses, threats to personal safety, or political conflicts~\cite{translation1, translation2, translation3, translation4, hsbcbank}. Thus, assuring the robustness of machine translation software is an important endeavor. 

Yet testing machine translation software is extremely challenging. First, different from traditional software whose logic is encoded in source code, machine translation software is based on complex neural networks with millions of parameters. Therefore, testing techniques for traditional software, which are mostly code-based, are ineffective. Second, the line of recent research on testing artificial intelligence (AI) software~\cite{Pei17SOSP, Goodfellow15ICLR, Jia17EMNLP, Mudrakarta18ACL, Alzantot18EMNLP, Iyyer18NAACL, Li19NDSS} focuses on tasks with much simpler output formats---for example, testing image classifiers, which output class labels given an image. However, as one of the most difficult natural language processing (NLP) tasks, the output of machine translation systems (i.e., translated sentences) is significantly more complex. Because they are not structured to handle such complex outputs, when applied to NMT models, typical AI testing approaches almost solely find "illegal" inputs, such as sentences with syntax errors or obvious misspellings that are unlikely given as input. Yet these errors are not the problematic ones in practice; as reported by WeChat, a messenger app with over one billion monthly active users, its embedded NMT model can return inferior results even when the input sentences are syntactically correct~\cite{Zheng18Arxiv}. Due to the difficulty of building an effective, automated approach to evaluate the correctness of translation, current approaches for testing machine translation software have many shortcomings.  


Approaches that try to address these aforementioned problems still have their own deficiencies---namely, the inability to detect grammatical errors and the lack of real-world test cases. Current testing procedures for machine translation software typically involve three steps \cite{Zheng18Arxiv}: (1) collecting bilingual sentence pairs\footnote{By a sentence pair, we refer to a source sentence and its corresponding target sentence.} and splitting them into training, validation, and testing data; (2) calculating translation quality scores (e.g., BLEU~\cite{Papineni02BLEU} and  ROUGE~\cite{Lin04ROUGE}) of the trained NMT model on the testing data; and (3) comparing the scores with predefined thresholds to determine whether the test cases pass. However, testing based on a threshold score like BLEU, which is a measurement of the overlap between n-grams of the target and reference, can easily overlook grammatical errors.  Additionally, the calculation of translation quality scores (e.g., BLEU) requires bilingual sentence pairs as input, which need to be manually constructed beforehand. To test with real-world user input outside of the training set, extensive manual effort is needed for ground-truth translations. Thus, an effective and efficient testing methodology that can automatically detect errors\footnote{By a translation error, we refer to mistranslation of some parts of a source sentence. The translated sentence (i.e., target sentence) containing translation error(s) is regarded as a buggy sentence. We use "error in the target sentence" and "error in the sentence pair" interchangeably in this paper.} in machine translation software is in high demand. 

\begin{figure*}[th]
\centering{} 
\includegraphics[scale=0.86]{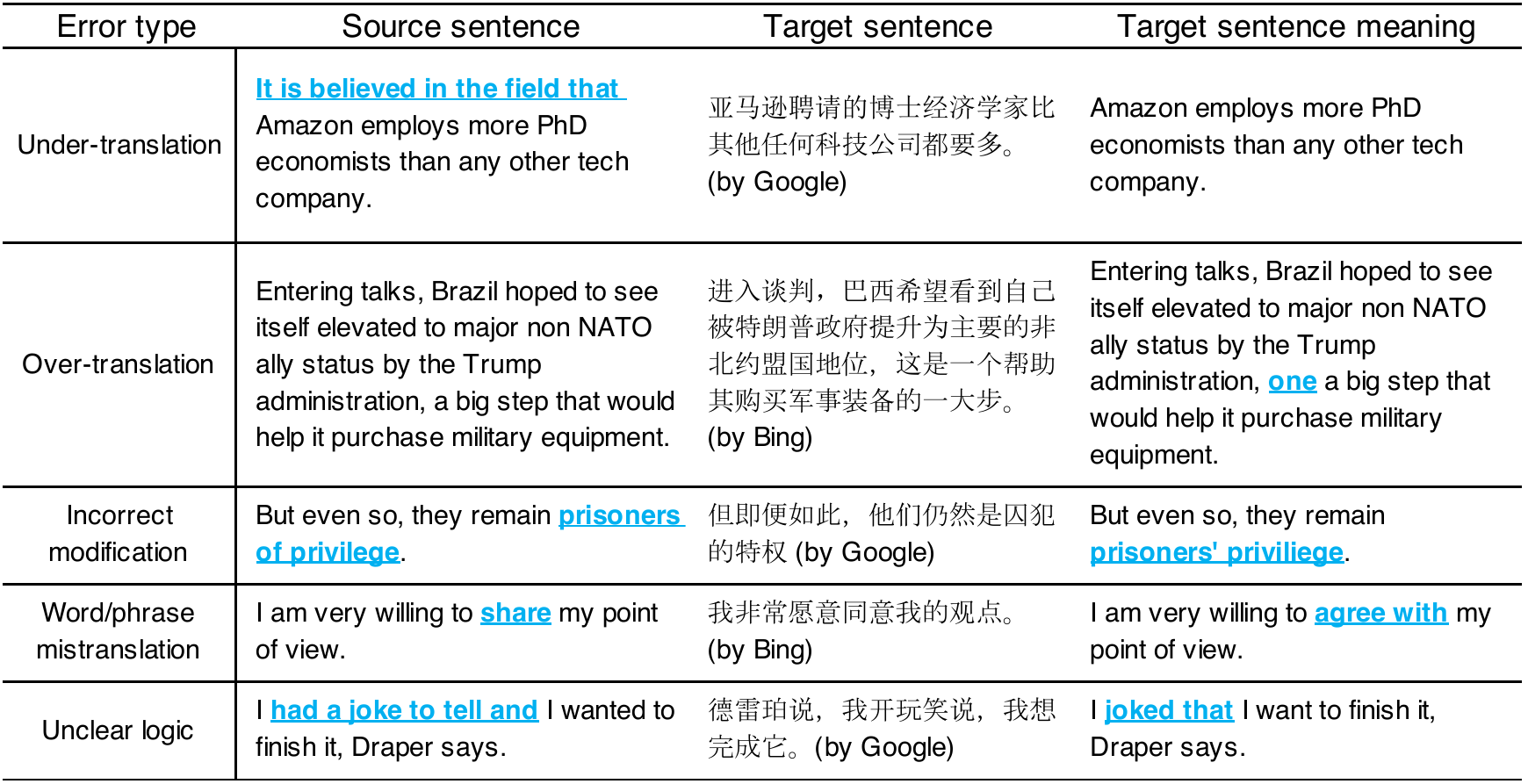}
\caption{Examples of translation errors (English-to-Chinese) detected by SIT.}
\label{fig:bugexample}
\end{figure*}

To address the above problems, we introduce structure-invariant testing (SIT), a novel, widely-applicable methodology for validating machine translation software. The key insight
is that similar source sentences---e.g. sentences that differ by a single word---typically have translation results of similar sentence structures. For example, Fig.~\ref{fig:sitinsight} shows three similar source sentences in English and their target sentences in Chinese. The first two translations are correct, while the third is not. We can observe that the structure of the third sentence in Chinese significantly differs from those of the other two. For each source sentence, SIT (1) generates a list of its similar sentences by modifying a single word in the source sentence via NLP techniques (i.e., BERT~\cite{Devlin18Bert}); (2) feeds all the sentences to the software under test to obtain their translations; (3) uses specialized data structures (i.e., constituency parse tree and dependency parse tree) to represent the syntax structure of each of the translated sentences; and (4) compares the structures of the translated sentences. If a large difference exists between the structures of the translated original and any of the translated modified sentences, we report the modified sentence pair along with the original sentence pair as potential errors.

We apply SIT to test Google Translate and Bing Microsoft Translator with 200 source sentences crawled from the Web as input. SIT successfully found 64 buggy issues (defined in Section~\ref{sec:method}) in Google Translate and 70 buggy issues in Bing Microsoft Translator with high accuracy (i.e., 69.5\% and 70\% top-1 accuracy respectively). The reported errors\footnote{https://github.com/PinjiaHe/StructureInvariantTesting} are diverse, including under-translation, over-translation, incorrect modification, word/phrase mistranslation, and unclear logic, none of which could be detected by the widely-used metrics BLEU and ROUGE. Examples of different translation errors are illustrated in Fig.~\ref{fig:bugexample}. The source code and datasets are also released for reuse. Note that our results were \emph{w.r.t.} the snapshots of Google Translate and Bing Microsoft Translator when we performed our testing.  After releasing our results dataset in July 2019, we notice that some of the reported translation errors have recently been addressed.

This paper makes the following main contributions:
\begin{itemize}
  \item It introduces structure-invariant testing (SIT), a novel, widely applicable methodology for validating machine translation software;
  \item It describes a practical implementation of SIT by adapting BERT~\cite{Devlin18Bert} to generate similar sentences and leveraging syntax parsers to represent sentence structures;
  \item It presents the evaluation of SIT using only 200 source sentences crawled from the Web to successfully find 64 buggy issues in Google Translate and 70 buggy issues in Bing Microsoft Translator with high accuracy; and
  \item It discusses the diverse error categories found by SIT, of which none could be found by state-of-the-art metrics.
  
\end{itemize}

\begin{figure*}[t]
\centering{}
 \includegraphics[scale=0.54]{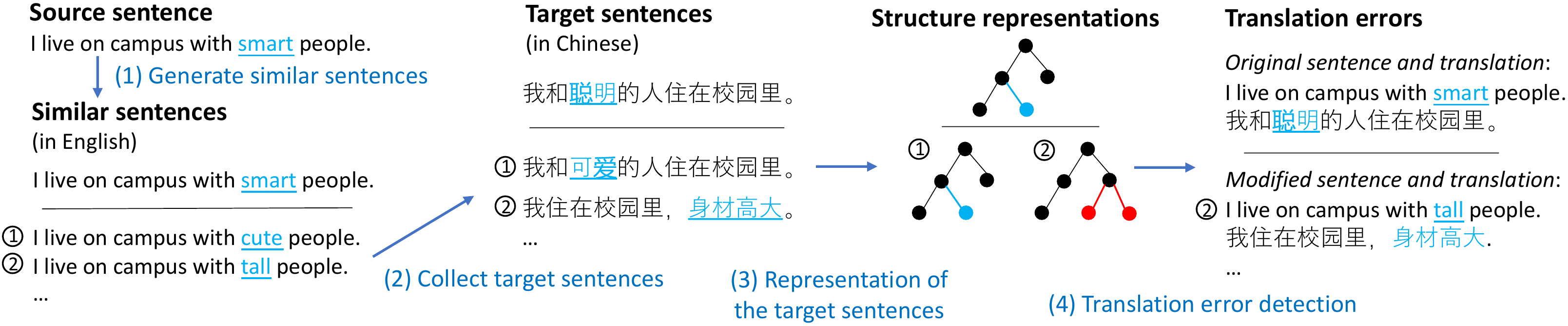} 
\caption{Overview of SIT.}
\label{fig:overview}
\end{figure*}

\section{A Real-World Example }\label{sec:motivating}

Tom planned to take his son David, who is 14 years old, to the Zurich Zoo. Before their zoo visit, he checked the zoo's website\footnote{https://www.zoo.ch/de/zoobesuch/tickets-preise} about purchasing tickets and saw the following German sentence:


\begin{center}
  \includegraphics[scale=0.42]{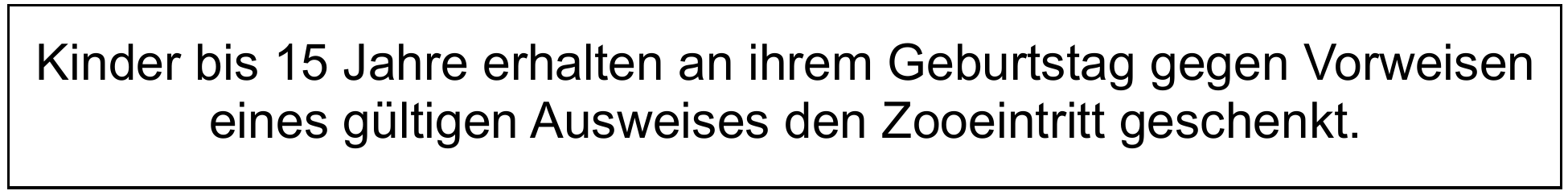}  
\end{center}

Tom is from the United States, and he does not understand German. To figure out its meaning, Tom used Google Translate, a popular translation service powered by NMT models~\cite{Wu16Arxiv}. Google Translate returned the following English sentence:

\begin{center}
    \includegraphics[scale=0.42]{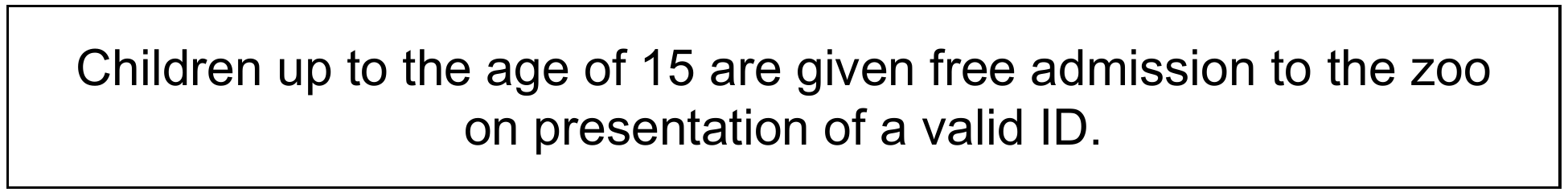}
\end{center}


However, David was denied free entry by the zoo staff even with a valid ID. They found out that they had misunderstood the zoo's regulation because of the incorrect translation returned by Google Translate. The correct translation should be:

\begin{center}
 \includegraphics[scale=0.42]{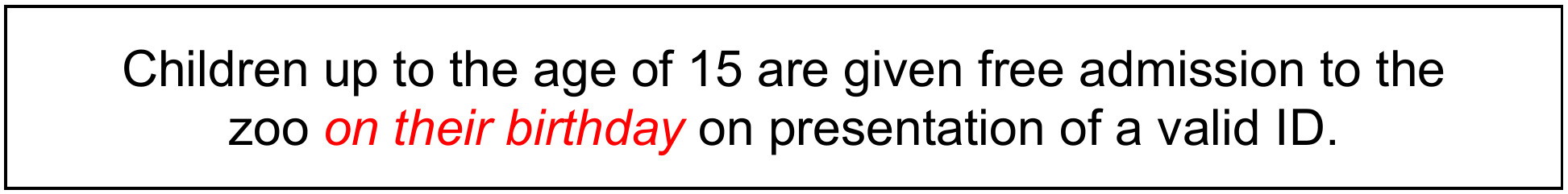}    
\end{center}

This is a real translation error that led to a confusing, unpleasant experience. Translation errors could also cause extremely serious consequences~\cite{translation2, hsbcbank, translation3, translation4}. For example, a Palestinian man was arrested by Israeli police for a post saying "good morning," which Facebook's machine translation service erroneously translated as "attack them" in Hebrew and "hurt them" in English~\cite{translation3, translation4}. This demonstrates both the widespread reliance on machine translation software and the potential negative effects when it fails. To enhance the reliability of machine translation software, this paper introduces a general validation approach called structure-invariant testing, which automatically and accurately detects translation errors without oracles.

  \section{Approach and Implementation}\label{sec:method}

This section introduces structure-invariant testing (SIT) and describes our implementation. The input of SIT is a list of unlabeled, monolingual sentences, while its output is a list of suspicious \textit{issues}. For each original sentence, SIT reports either 0 (i.e., no buggy sentence is found) or 1 \textit{issue} (i.e., at least 1 buggy sentence is found). Each \textit{issue} contains: (1) the original source sentence and its translation; and (2) top-k farthest\footnote{the distance metric here is between the structures of the original sentence translation and the modified sentence translations} generated source sentences and their translations.  The original sentence pair is reported for the following reasons: (1) seeing how the original sentence was modified may help the user understand why the translation system made a mistake; (2) the error may actually lie in the translation of the original sentence.

Fig.~\ref{fig:overview} illustrates the overview of SIT. In this figure, we use one source sentence as input for simplicity and clarity.
The key insight of SIT is that similar source sentences often have target sentences of similar syntactic structure. 
Derived from this insight, SIT carries out the following four steps:

\begin{enumerate}
  \item \textit{Generating similar sentences.}\enspace For each source sentence, we generate a list of its similar sentences by modifying a single word in the sentence.
  \item \textit{Collecting target sentences.}\enspace We feed the original and the generated similar sentences to the machine translation system under test and collect their target sentences.
  \item \textit{Representing target sentence structures.}\enspace All the target sentences are encoded as data structures specialized for natural language processing.
  \item \textit{Detecting translation errors.}\enspace The structures of the translated modified sentences are compared to the structure of the translated original sentence. If there is a large difference between the structures, SIT reports a potential error. 
\end{enumerate}



\subsection{Generating Similar Sentences}
In order to test for structural invariance, we must compare two sentences that have the same syntactic structure but differ in at least one token.
We have found that, given an input sentence, changing one word in the sentence at a time under certain constraints effectively produces a set of structurally identical and semantically similar sentences.

\begin{figure}[t]
\centering{}
 \includegraphics[scale=0.65]{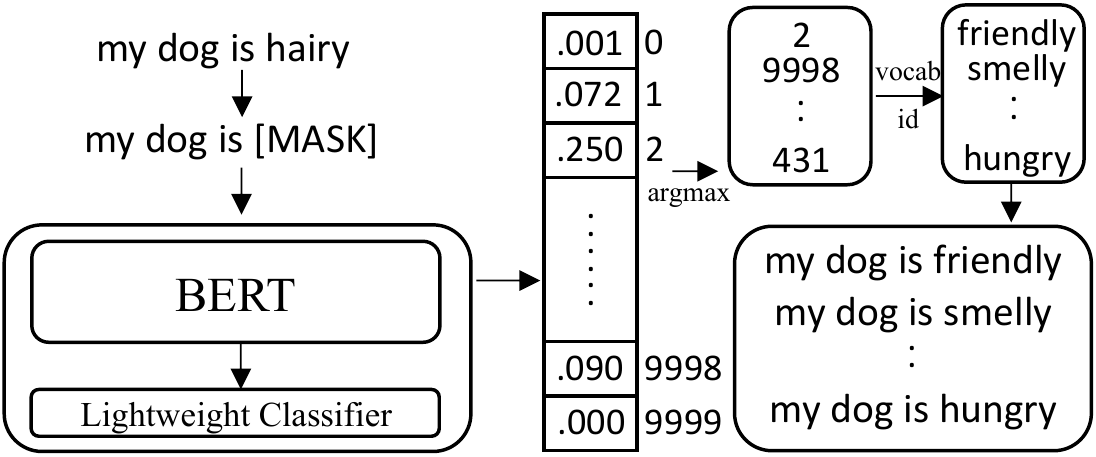} 
\caption{Similar sentence generation process.}
\vspace{-5ex}
\label{fig:similarsent}
\end{figure}

Explicitly, the approach we take modifies a single token in an input sentence, replacing it with another token of the same part of speech (POS),\footnote{https://en.wikipedia.org/wiki/Part\_of\_speech} to produce an alternate sentence. For example, we will mask "hairy" in the source sentence in Fig.~\ref{fig:similarsent} and replace it with the top-k most similar tokens to generate k similar sentences. We do this for every candidate token in the sentence; for the sake of simplicity and to avoid grammatically strange or incorrect sentences, we only use nouns and adjectives as candidate tokens. 

Now we discuss the problem of selecting replacement tokens. Perhaps the simplest algorithm for selecting a set of replacement tokens would involve using word embeddings~\cite{Mikolov13Wordembedding}. One could choose words that have high vector similarity with and identical POS tags to a given token in the original sentence as replacements in the modified sentences. However, since word embeddings have the same value regardless of context, this approach often produces sentences that would not occur in common language. For example, the word "fork" might have high vector similarity with and the same POS tag as the word "plate." However, while the sentence "He came to a fork in the road" makes sense, the sentence "He came to a plate in the road" does not.

Rather, we want a model that considers the surrounding words and comes up with a set of replacements that, when inserted, create realistic sentences. A model that does just this is the masked language model (MLM)~\cite{MikolovICLRW13}, inspired by the Cloze task ~\cite{TaylorCloze}. The input to an MLM is a piece of text with a single token masked (i.e., deleted from the sentence and replaced with a special indicator token). The job of the model is then to predict the token in that position given the context. This method forces the model to learn the dependencies between different tokens. Since there are a number of different contexts a single word can fit in, this model, in a sense, allows for a single token to have multiple representations. We therefore get a set of replacement tokens that are context dependent. While the predicted tokens are not guaranteed to have the same meaning as the original token, if the MLM is well trained, it is highly likely that the sentence with the new, predicted token is both syntactically correct and meaningful.

An example of the sentence generation process is demonstrated in Fig.~\ref{fig:similarsent}. For our implementation, we use BERT~\cite{Devlin18Bert}, which is a state-of-the-art language representation model recently proposed by Google. The out-of-box BERT model provides pre-trained language representations that can be fine-tuned by adding an additional lightweight softmax classification layer to create models for a wide range of language-related tasks, such as masked language modelling.  BERT was trained on a huge amount of data---a concatenation of BooksCorpus (800M words) and English Wikipedia (2,500M words)---with the masked language task being one of two main tasks used for training. Thus, we believe that BERT fits this aspect of our approach well.

\subsection{Collecting Target Sentences}

Once we have generated a list of similar sentences from our original sentence, the next step is to input all the source sentences to the machine translation software under test and collect the corresponding translation results (i.e., target sentences). We subsequently analyze the results to find errors.
We use Google's and Bing's machine translation systems as test systems for our experiment. To obtain translation results, we invoke the APIs provided by Google Translate and Bing Microsoft Translator, which return identical results as their Web interfaces~\cite{googletranslate, bingtranslator}.

\subsection{Representations of the Target Sentences}

Next we must model the target sentences obtained from the translation system under test as this allows us to compare structures to detect errors. Choosing the structure with which to represent our sentences will affect our ability to perform meaningful comparisons. We ultimately want a representation that precisely models the structure of a sentence while offering fast comparison between two values.

The simplest and fastest approach is to compare sentences in their raw form: as strings. Indeed, we test this method and performance is reasonable. However, there are many scenarios in which this method falls short.  For example, the prepositional phrase "on Friday" in the sentence "On Friday, we went to the movies" can also be placed on the end of the sentence as follows: "We went to the movies on Friday." The sentences are interchangeable but a metric such as character edit distance
would indicate a large difference between the strings. 
Syntax parsing overcomes the above issue. With a syntax parser, we can model the syntactic structure of a string and the relationship between words or groups of words. For example, if parsing is done correctly, our two sample sentences above should have identical representations in terms of relation values and parse structure. 




\begin{figure}[t]
\centering{}
 \includegraphics[scale=0.56]{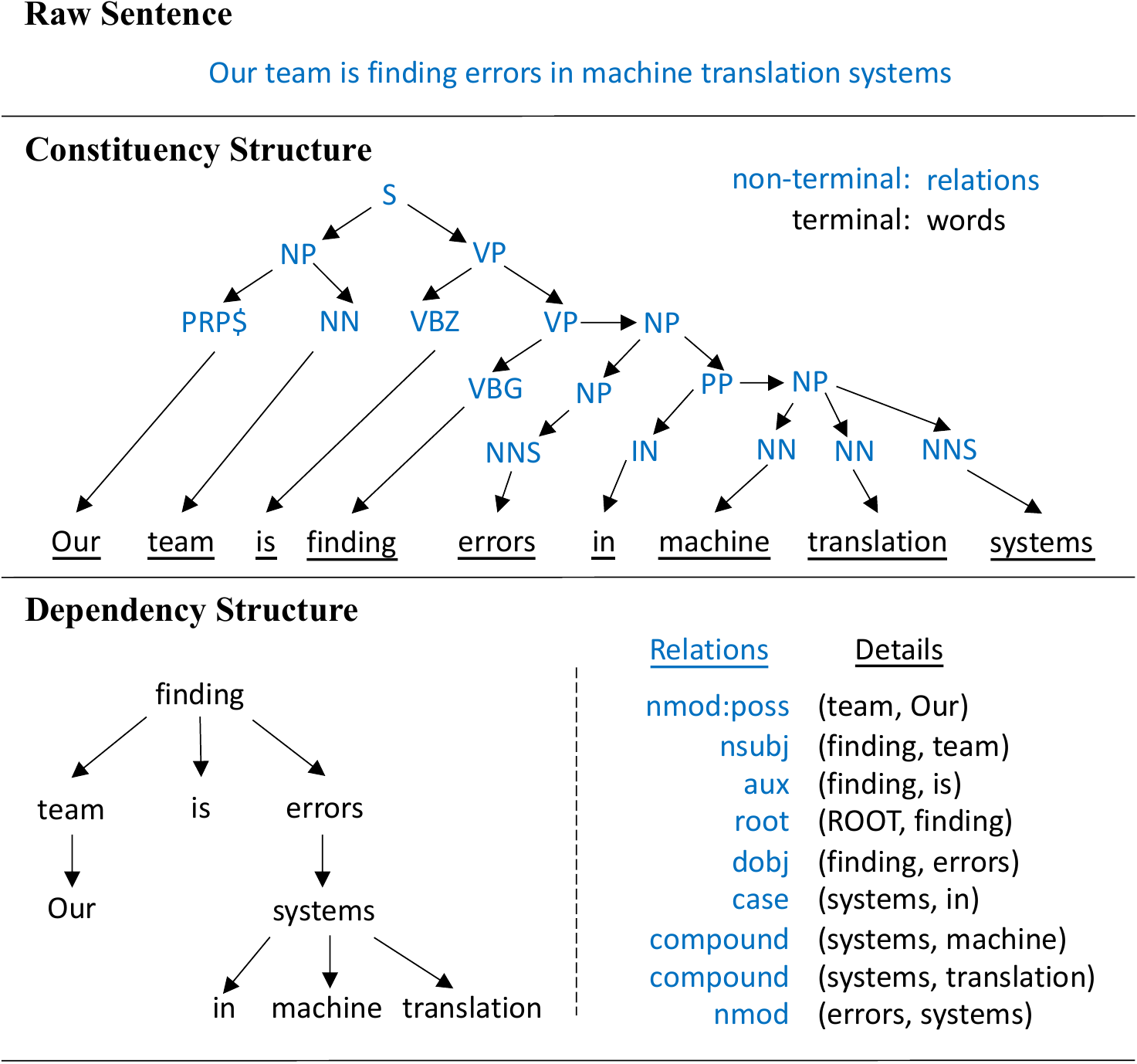} 
\caption{Representing sentence structures; both dependency \& constituency relations can be displayed as trees.}
\label{fig:sentrep}
\end{figure}

\subsubsection{Raw Target Sentence}
For this method, we leave our target sentence in its original format, i.e., as a string. In most cases, we may expect that editing a single token in a sentence in one language would lead to the change of a single token in the translated sentence. Given the syntactic role of the replacement token is the same, this would ideally happen in all machine translation systems. However, this is not always the case in practice as prepositional phrases, modifiers, and other constituents can often be placed in different locations by the translation system and produce a semantically-equivalent, grammatically correct sentence. Nonetheless, this method serves as a good baseline.

\subsubsection{Constituency Parse Tree}
Constituency parsing is one method for deriving the syntactic structure of a string. It generates a set of constituency relations, which show how a word or group of words form different units within a sentence. This set of relations is particularly useful for SIT because it will reflect changes to the type of phrases in a sentence. For example, while a prepositional phrase can be placed in multiple locations to produce a sentence with the same meaning, the set of constituency relations will remain unchanged.
Constituency relations can be visualized as a tree, as shown in Fig.~\ref{fig:sentrep}. A constituency parse tree is an ordered, rooted tree where non-terminal nodes are the constituent relations and terminal nodes are the words.
Formally, in constituency parsing, a sentence is broken down into its constituent parts according to the phrase structure rules ~\cite{ChomskyCFG} outlined by a given context-free grammar. 
For our experiments, we use the shift-reduce constituency parser by Zhu \emph{et al.}~\cite{SRConstParsing} and implemented in Stanford's CoreNLP library~\cite{stanfordcorenlp}. It can parse about 50 sentences per second.


\subsubsection{Dependency Parse Tree}
Dependency parsing likewise derives the syntactic structure of a string. However, the set of relations produced describe the direct relationships between words rather than how words constitute a sentence. This set of relations gives us different insights about structure and is intuitively useful for SIT because it will reflect changes between how words interact. 
Much progress has been made over the past 15 years on dependency parsing. Speed and accuracy increased dramatically with the introduction of neural network based parsers~\cite{Chen14EMNLP}. As with shift-reduce constituency parsers, neural network based dependency parsers use a stack-like system where transitions are chosen using a classifier. The classifier in this case is a neural network, likewise trained on annotated tree banks. For our implementation, we use the most recent neural network based parsers made available by Stanford CoreNLP, which can parse about 100 sentences per second. We use the Universal Dependencies as our annotation scheme, which has evolved based off the Stanford Dependencies~\cite{USDependencies}. 





\subsection{Translation Error Detection via Structure Comparison}
Finally, in order to find translation errors, we search for structural variation by comparing sentence representations.
Whether sentences are modelled as raw strings, word embeddings, or parse trees, there are a number of different metrics for calculating the distance between two values. These metrics tend to be quite domain specific and might have low correlation with each other, making the choice of metric incredibly important.  
For example, a metric such as Word Mover's Distance~\cite{WMD} would give us a distance of 0 between the two sentences "He went to the store" and "Store he the went to" while character edit distance would give a distance of 14. 
We explore several different metrics for evaluating the distance between sentences: character (Levenshtein) edit distance, constituency set difference, and dependency set difference.


\subsubsection{Levenshtein Distance between Raw Sentences}

The Levenshtein distance~\cite{Lev}, sometimes more generally referred to as the "edit distance," compares two strings and determines how closely they match each other by calculating the minimum number of character edits (deletions, insertions, and substitutions) needed to transform one string into the other.  
While the method may not demonstrate syntactic similarity between sentences well, it exploits the expectation that editing a single token in a sentence in one language will often lead to the change of only a single token in the translated sentence. Therefore, the Levenshtein distance serves as a good baseline metric.

\subsubsection{Relation Distance between Constituency Parse Trees}
To evaluate the distance between two sets of constituency relations, we calculate the distance between two lists of constituency grammars as the sum of absolute difference in the count of each phrasal type, which gives us a basic understanding of how a sentence has changed after modification.
The motivation behind this heuristic is that the constituents of a sentence should stay the same between two sentences where only a single token of the same part of speech differs. In a robust machine translation system, this should be reflected in the target sentences as well.

\subsubsection{Relation Distance between Dependency Parse Trees}
Similarly, for calculating the distance between two lists of dependencies, we sum the absolute difference in the number of each type of dependency relations.
Again, the motivation is that the relationships between words will ideally remain unchanged when a single token is replaced. Therefore, a change in the set is reasonable indication that structural invariance has been violated and presumably there is a translation error. 

\subsubsection{Distance Thresholding}
Using one of the above metrics, we calculate the distance between the original target sentence and the generated target sentences. We must then decide whether a modified target sentence is far enough from the its corresponding original target sentence to indicate the presence of a translation error. 
To do this, we first filter based on a distance threshold, only keeping sentences that are farther from the original sentence than the chosen threshold. Then, for a given original target sentence, we report the top-\textit{k} (\textit{k} also being a chosen parameter) farthest modified target sentences. 
We leave the distance threshold as a manual parameter since the user may prioritize minimizing false positive reports or minimizing false negative reports depending on their goal. In Section~\ref{sec:threshold}, we show tradeoffs for different threshold values. 
For each original sentence, an issue will be reported if at least one translated generated sentence is considered buggy. 

\section{Evaluation}\label{sec:exper}

In this section, we evaluate our approach by applying it to Google Translate and Bing Microsoft Translator with real-world unlabeled sentences crawled from the Web. Our main research questions are:

\begin{itemize}
  \item RQ1: How effective is the approach at finding buggy translations in machine translation software?
  \item RQ2: What kinds of translation errors can our approach find?
  \item RQ3: How efficient is the approach?  
  \item RQ4: How do we select the distance threshold in practice?
\end{itemize}


\subsection{Experimental Setup}
To verify the results of SIT, we manually inspect each issue reported and collectively decide: (1) whether the issue contains buggy sentences; and (2) if yes, what kind of translation errors it contains. All experiments are run on a Linux workstation with 6 Core Intel Core i7-8700 3.2GHz Processor, 16GB DDR4 2666MHz Memory, and GeForce GTX 1070 GPU. The Linux workstation is running 64-bit Ubuntu 18.04.02 with Linux kernel 4.25.0. 

\subsection{Dataset}
Typically, to test a machine translation system, developers can adopt SIT with any source sentence as input. Thus, to evaluate the effectiveness of our approach, we collect real-world source sentences from the Web. Specifically, input sentences are extracted from CNN\footnote{https://edition.cnn.com/} (Cable News Network) articles in two categories: politics and business. The datasets are collected from two categories of articles because we intend to evaluate whether SIT consistently performs well on sentences of different semantic context.

For each category, we crawled the 10 latest articles, extracted their main text contents, and split them into a list of sentences. Then, we randomly select 100 sentences from each sentence list as the experimental datasets (200 in total). In this process, sentences that contain more than 35 words are filtered because we intend to demonstrate that machine translation software can return inferior results even for relatively short, simple sentences. The details of the collected datasets are illustrated in Table~\ref{tab:datasets}.

\begin{table}[h]
\centering{}
\caption{Statistics of input sentences for evaluation. Each corpus contains 100 sentences.}
\includegraphics[scale=0.45]{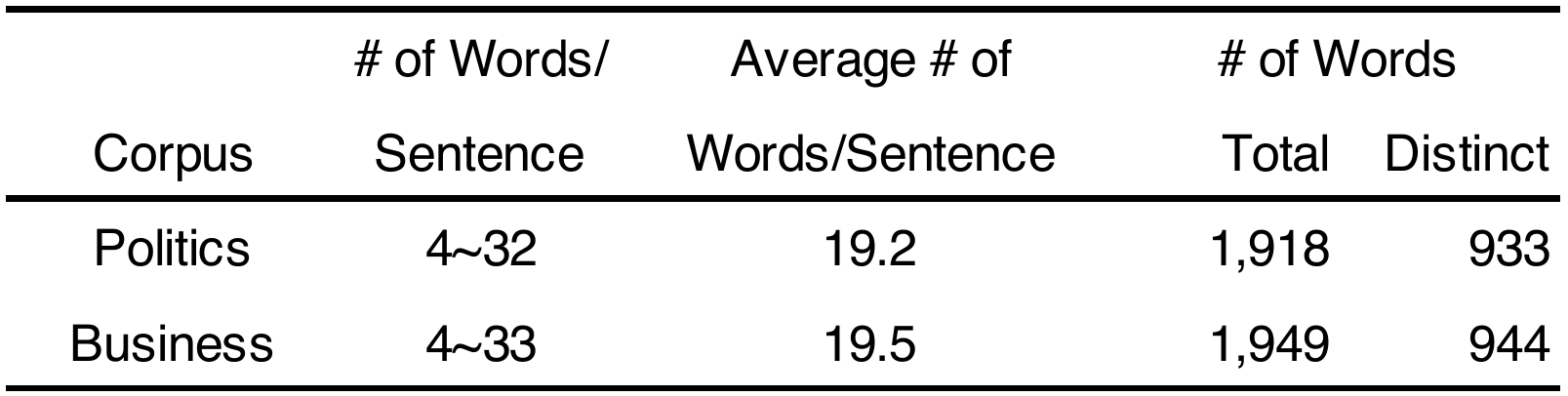}
\label{tab:datasets}
\end{table}

\subsection{The Effectiveness of SIT}

Our approach aims to automatically find translation errors using unlabeled sentences and report them to developers. Thus, the effectiveness of the approach lies in two aspects: (1) how accurate are the reported results; and (2) how many buggy sentences can SIT find? In this section, we evaluate both aspects by applying SIT to test Google Translate and Bing Microsoft Translator using the datasets illustrated in Table~\ref{tab:datasets}.

\subsubsection{Evaluation Metric}
The output of SIT is a list of \textit{issues}, each containing (1) an original source sentence and its translation; (2) the top-$k$ reported generated sentences and their translations (i.e. the $k$ farthest translations from the source sentence translation). Here we define top-$k$ accuracy as the percentage of reported issues where at least one of the top-k reported sentences or the original sentence contains an error. We use this as our accuracy metric for SIT. Explicitly, if there is a buggy sentence in the top-$k$ generated sentences of issue $i$, we consider the issue to be accurate and set $buggy(i, k)$ to true; else we set $buggy(i, k)$ to false. If the original sentence is buggy and was reported as an issue, then we also set $buggy(i, k)$ to true. Given a list of issues $I$, its top-$k$ accuracy is calculated as:



\begin{equation}
\label{equ:topk}
\textrm{Accuracy}_k = \frac{\sum_{i \in I} \mathds{1}\{buggy(i, k)\}}{|I|},
\end{equation}
where $|I|$ is the number of the issues returned by SIT.

\subsubsection{Results}\label{sec:accuracy}


\begin{table}[t]
\centering{}
\caption{Top-k accuracy of SIT.}
\includegraphics[scale=0.48]{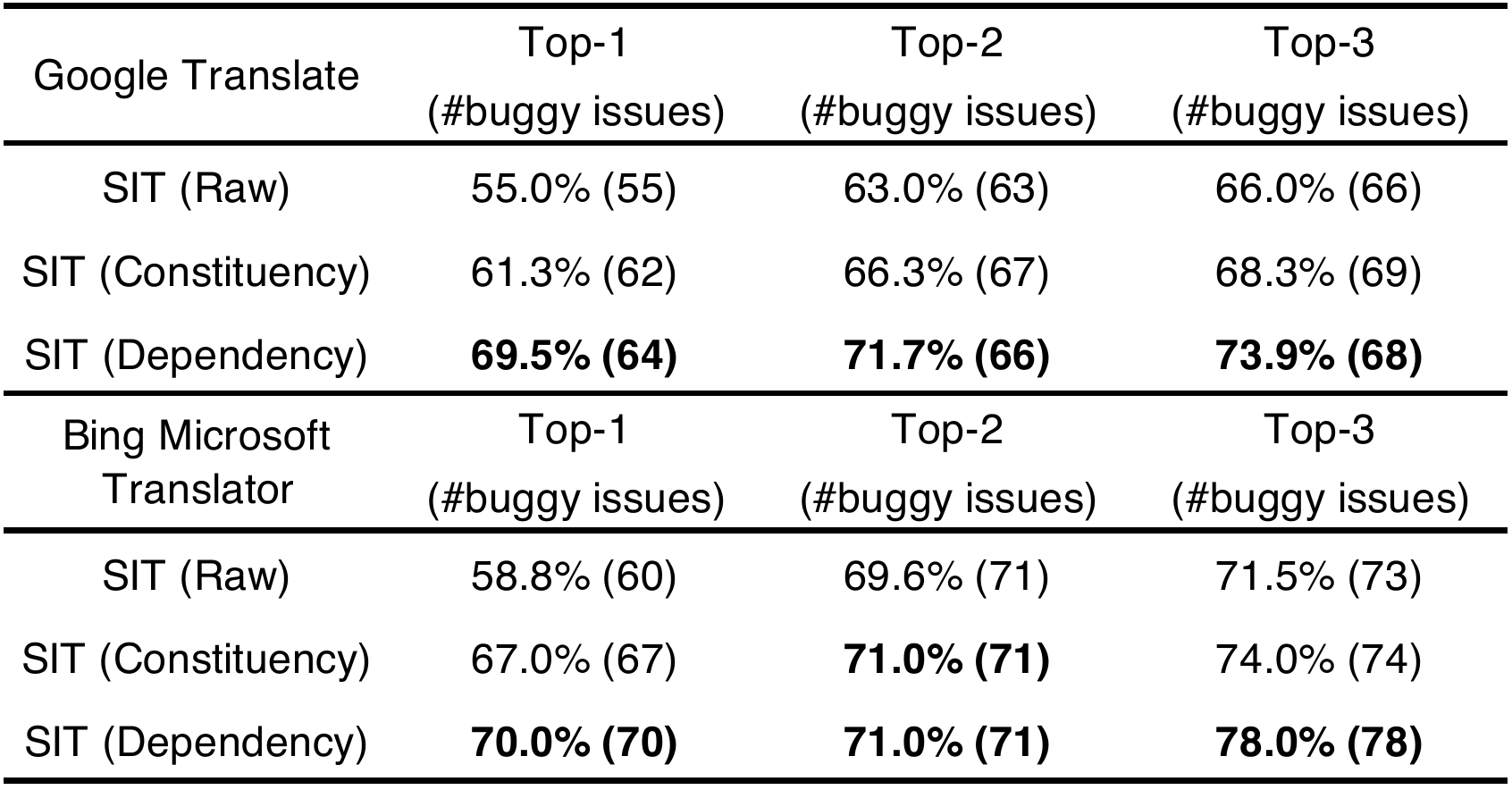}
\label{tab:accuracytop0}
\end{table}

\textbf{Top-k accuracy.} The results are summarized in Table~\ref{tab:accuracytop0}. SIT (Raw), SIT (Constituency), and SIT (Dependency) are SIT implementations with raw sentence, constituency structure, and dependency structure as sentence structure representation, respectively. Each item in the table presents the top-k accuracy along with the number of buggy issues found. In subsequent discussions, for brevity, we refer SIT (Constituency) and SIT (Dependency) as SIT (Con) and SIT (Dep), respectively.


We observe that SIT (Con) and SIT (Dep) consistently perform better than SIT (Raw), which demonstrates the importance of the structure representation of sentences. The metric used in SIT (Raw), which is based only on the characters in the sentences, is brittle and subject to over and under report errors. For example, SIT (raw) may report sentences that are different in word level but similar in sentence structure, leading to false positives. SIT (Con) and SIT (Dep) achieve comparable performance in terms of both top-k accuracy and the number of reported buggy issues. In particular, when testing Bing Microsoft Translator, SIT (Dep) reports 100 suspicious issues. Among these issues, 70 of them contain translation errors in the first reported sentence or the original sentence, achieving 70\% top-1 accuracy. SIT (Dep) has the best performance on Top-1 accuracy for both Google Translate and Bing Microsoft Translator. It successfully finds 64 and 70 buggy issues with 69.5\% and 71\% top-1 accuracy, respectively. SIT (Dep) also achieves the highest top-3 accuracy (73.9\% and 78\%). Note that source sentences in the same issue only differ by one word. Thus, inspecting top-3 sentences will not cause more effort compared with inspecting top-1 sentences.



In addition, we study whether SIT can trigger new errors in the generated sentences. As illustrated in Table~\ref{tab:uniquebugs}, in the reported issues, 55 and 60 unique errors are found in the translation of original sentences by Google Translate and Bing Microsoft Translator respectively. Besides these errors, SIT finds 79 and 66 extra unique errors that are revealed only in the generated sentence pairs but not in the original. Thus, given its high top-k accuracy and lots of extra unique errors reported, we believe SIT is very useful in practice.

We did not compared SIT's accuracy with \cite{Zheng18Arxiv} and \cite{Zhou18ASWEC} because of the following reasons. SIT targets general mistranslation errors, while~\cite{Zheng18Arxiv} focuses on under-/over-translations. Thus, we did not empirically compare with it. In terms of error type and quantity, \cite{Zheng18Arxiv} can only find some under-/over-translation errors in original sentence translations, while SIT finds general errors in translations of both original sentences and their derived similar sentences. \cite{Zhou18ASWEC} requires input sentences with specialized structures and thus it cannot detect any errors using our datasets.





\begin{table}[t]
\centering{}
\caption{Number of unique errors. Top-k unique errors by SIT are errors only in generated sentences output by SIT (Dep).}
\includegraphics[scale=0.45]{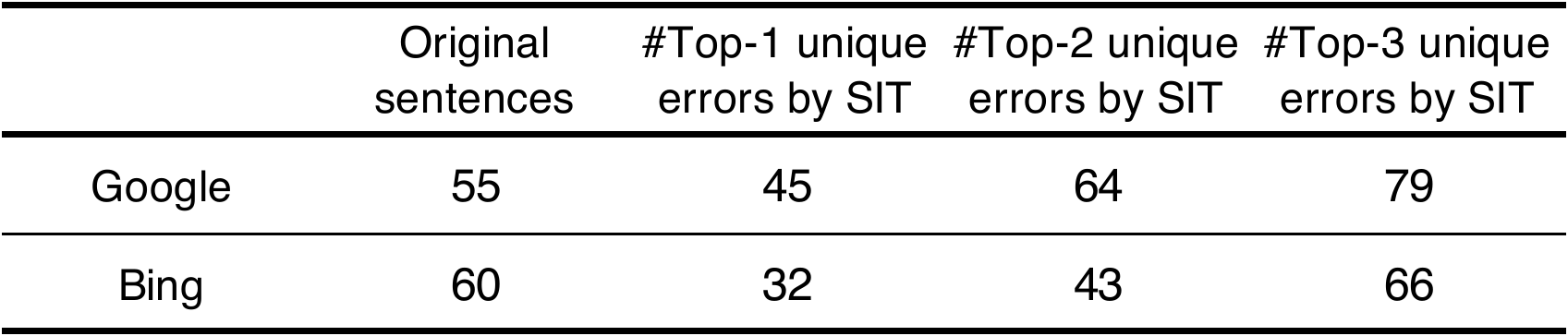}
\label{tab:uniquebugs}
\end{table}

\subsection{Translation Error Reported by SIT}
\begin{table}[t]
\centering{}
\caption{Number of sentences that have specific errors in each category SIT (Dep).}
\includegraphics[scale=0.45]{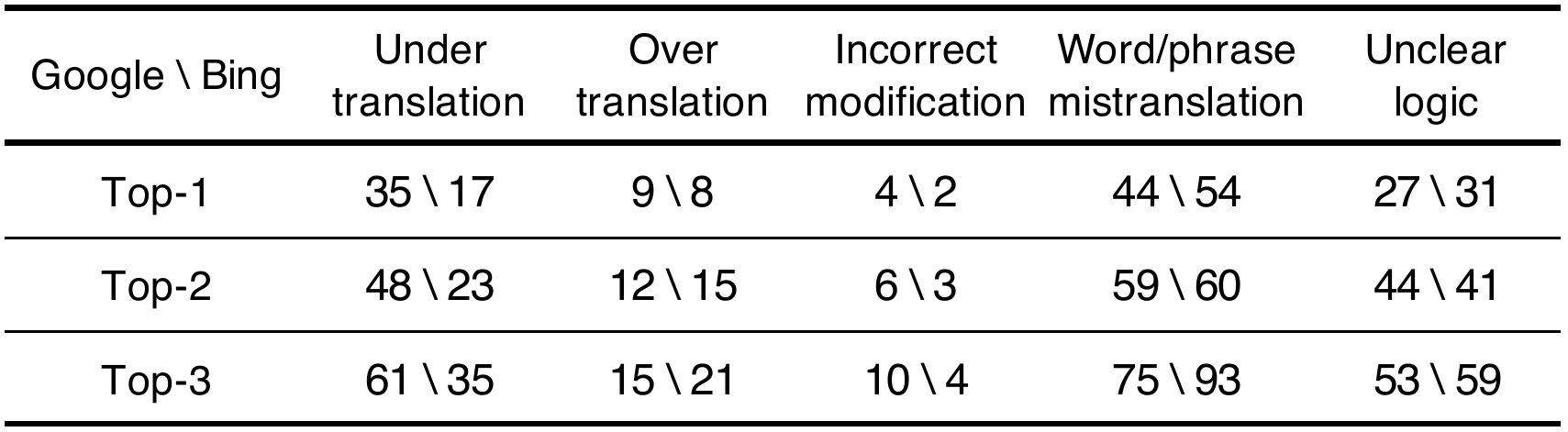}
\label{tab:bugtypesGoogle}
\end{table}


SIT is capable of finding translation errors of diverse kinds. In our experiments with Google Translate and Bing Microsoft Translator, we mainly find 5 kinds of translation errors: under-translation, over-translation, incorrect modification, word/phrase mistranslation, and unclear logic. The error types are derived from error classification methods for machine translation. Each of the five is a subset of lexical, syntactic, or semantic errors~\cite{Hsu14error}. We rename them in a more intuitive manner to aid the readers. To provide a glimpse of the diversity of the uncovered errors, this section highlights examples for all the 5 kinds of errors. Table~\ref{tab:bugtypesGoogle} presents the statistics of the translation errors SIT found. Under-translation, word/phrase mistranslation, and unclear logic account for most of the translation errors found by SIT.

\subsubsection{Under-Translation}

\begin{figure}[h]
\centering{} 
\includegraphics[scale=0.56]{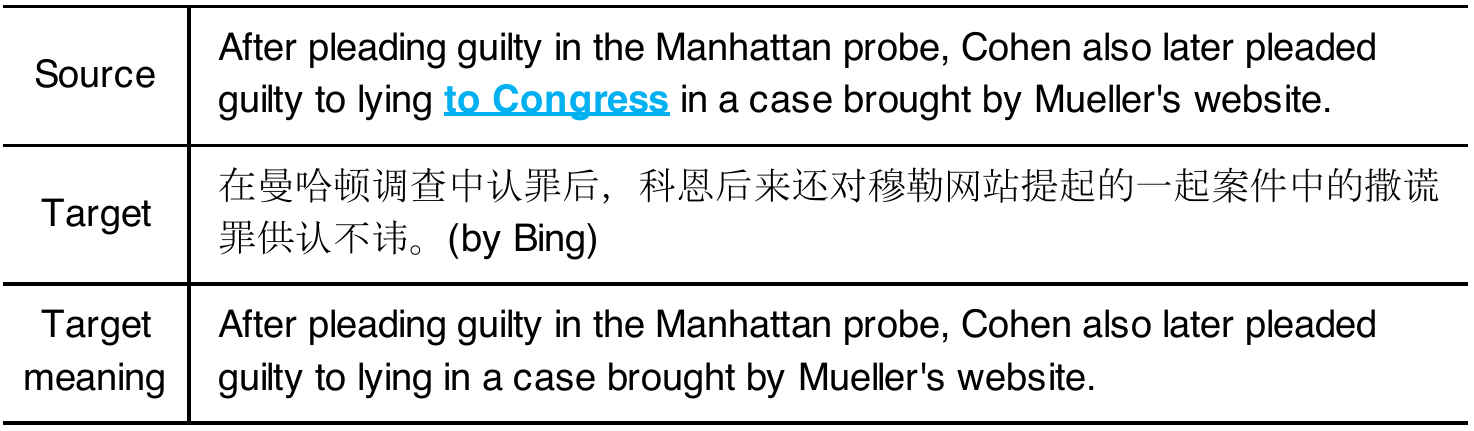}
\caption{Example of under-translation errors detected.}
\label{fig:undertranslation}
\end{figure}

If some words are mistakenly untranslated (i.e. do not appear in the translation), it is an under-translation error. Fig.~\ref{fig:undertranslation} presents a sentence pair that contains under-translation error. In this example, "to Congress" is mistakenly untranslated, which leads to target sentences of different semantic meaning. Specifically, "lying to Congress" is illegal while "lying" is just an inappropriate behavior. Likewise, the real-world example introduced in Section~\ref{sec:motivating} is caused by an under-translation error. 

\subsubsection{Over-Translation}

\begin{figure}[h]
\centering{} 
\includegraphics[scale=0.56]{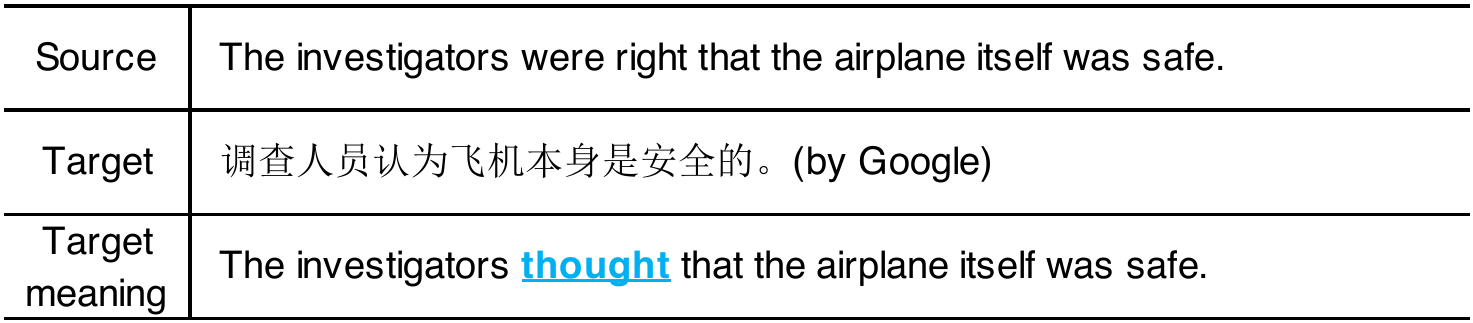}
\caption{Example of over-translation errors detected.}
\label{fig:overtranslation}
\end{figure}

If some words are unnecessarily translated multiple times or some words in the target sentence are not translated from any words in the source sentence, it is an over-translation error. In Fig.~\ref{fig:overtranslation}, "thought" in the target sentence is not translated from any words in the source sentence, so it is an over-translation error. Interestingly, we found that an over-translation error often happens along with some other kinds of errors. The example also contains an under-translation error because "were right" in the source sentence is mistakenly untranslated. In the second example in Fig.~\ref{fig:bugexample}, the word "a" is unnecessarily translated twice, which makes it an over-translation error.

\subsubsection{Incorrect Modification}

\begin{figure}[h]
\centering{} 
\includegraphics[scale=0.56]{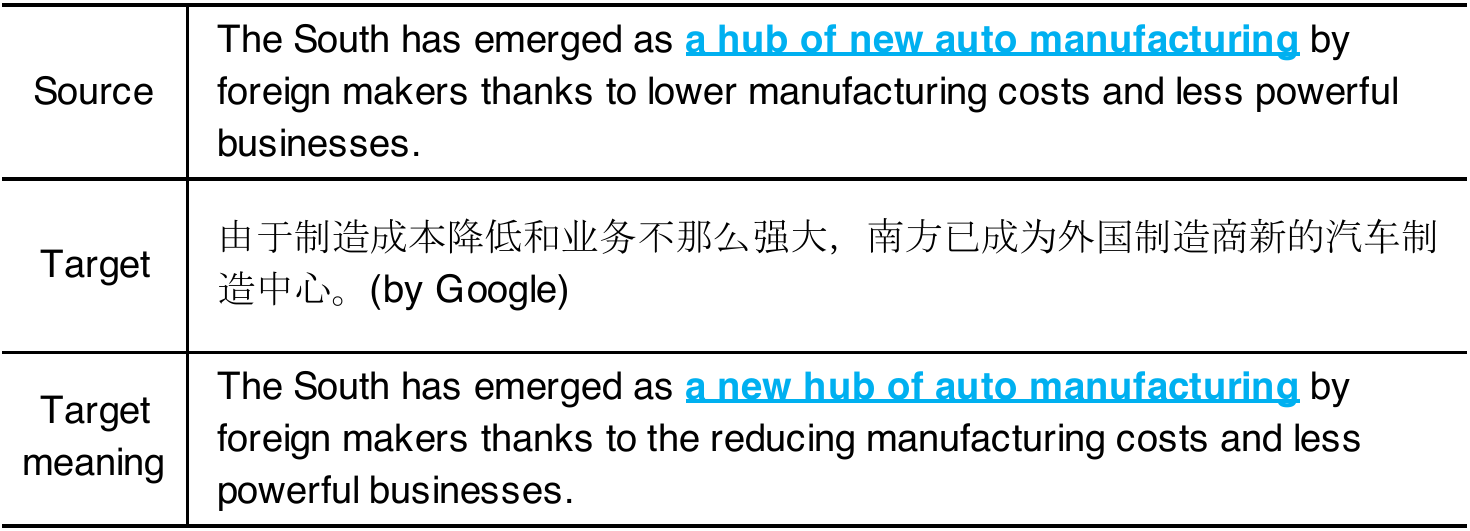}
\caption{Example of incorrect modification errors detected.}
\label{fig:incorrectmodification}
\end{figure}

If some modifiers modify the wrong element in the sentence, it is an incorrect modification error. In Fig.~\ref{fig:incorrectmodification}, the modifier "new" modifies "auto manufacturing" in the source sentence. However, Google Translate thinks that "new" should modify "hub." In Fig.~\ref{fig:bugexample}, the third example also shows an interesting incorrect modification error. In this example ("prisoners of privilege"), "privilege" modifies "prisoners" in the source sentence, while Google Translate thinks "prisoners" should modify "privilege." We think that in the training data of the NMT model, there are some phrases with the similar pattern: "\textit{A} of \textit{B}," where \textit{A} modifies \text{B}, which leads to an incorrect modification error in this scenario. Interestingly, the original source sentence that triggers this error is "But even so, they remain \textit{bastions of privilege}." In the original sentence, "bastions" modifies "privilege," which fits the supposed archetype. As we might expect, this sentence is correctly translated by Google Translate.

\subsubsection{Word/phrase Mistranslation}

\begin{figure}[h]
\centering{} 
\includegraphics[scale=0.56]{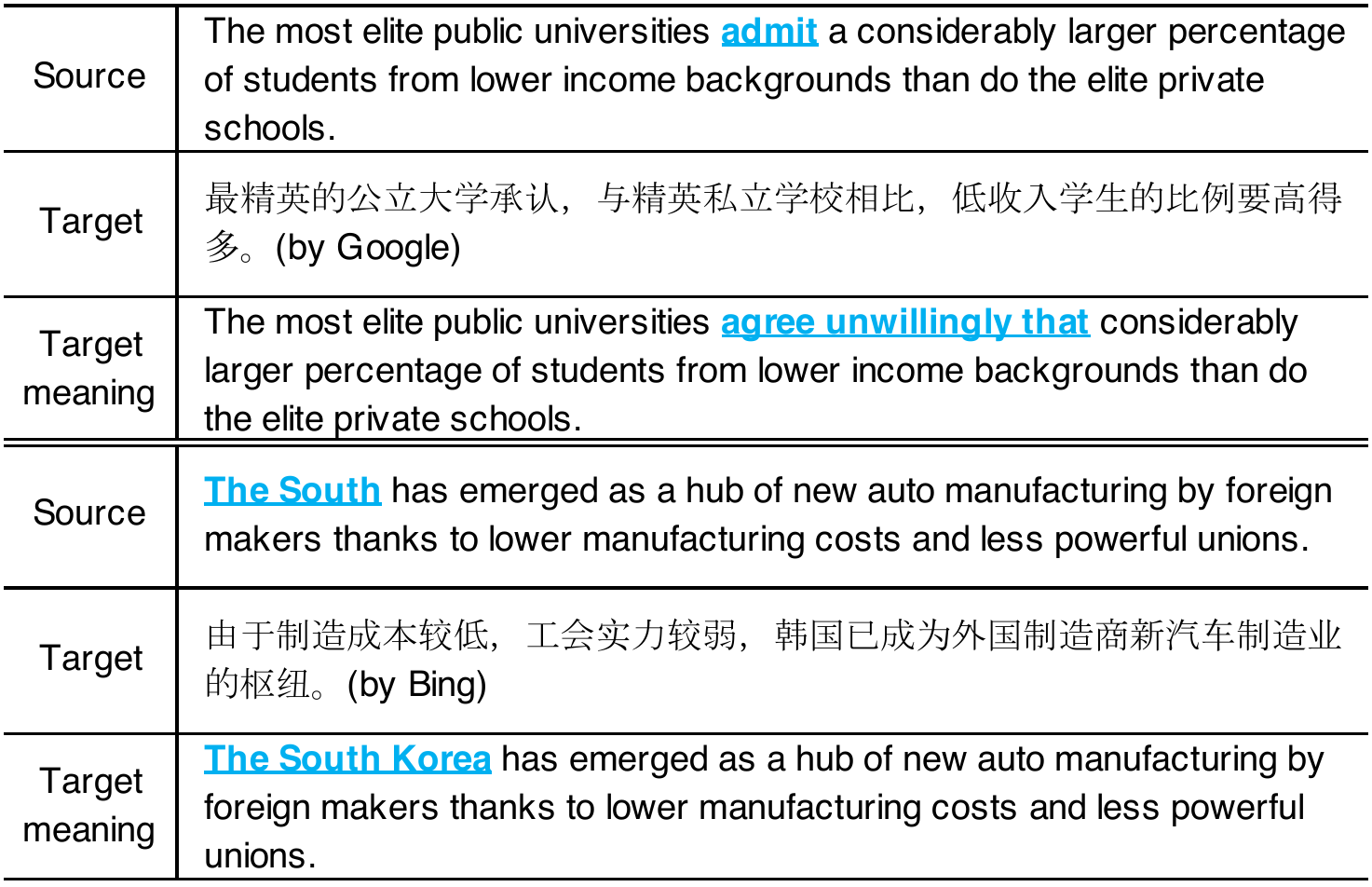}
\caption{Examples of word/phrase mistranslation errors detected.}
\label{fig:ambiguityofpolysemy}
\end{figure}

If some tokens or phrases are incorrectly translated in the target sentence, it is a word/phrase mistranslation error. Fig.~\ref{fig:ambiguityofpolysemy} presents two main sub-categories of this kind of error: (1) ambiguity of polysemy and (2) wrong translation. 

\vspace*{3pt}
\noindent
\textbf{Ambiguity of polysemy.}\quad Each token/phrase may have multiple correct translations. For example, admit means "allow somebody to join an organization" or "agree with something unwillingly." However, usually in a specific semantic context (e.g., a sentence), a token/phrase only has one correct translation. Modern translation software does not perform well on polysemy. In the first example in Fig.~\ref{fig:ambiguityofpolysemy}, Google Translate thinks the "admit" in the source sentence refers to "agree with something unwillingly," leading to a token/phrase mistranslation error.

\vspace*{3pt}
\noindent
\textbf{Wrong translation.}\quad A token/phrase could also be incorrectly translated to another meaning that seems semantically unrelated. For example, in the second example in Fig.~\ref{fig:ambiguityofpolysemy}, Bing Microsoft Translator thinks "South" refers to "South Korea," leading to a word/phrase mistranslation error.

\subsubsection{Unclear Logic}
\begin{figure}[h]
\centering{} 
\includegraphics[scale=0.56]{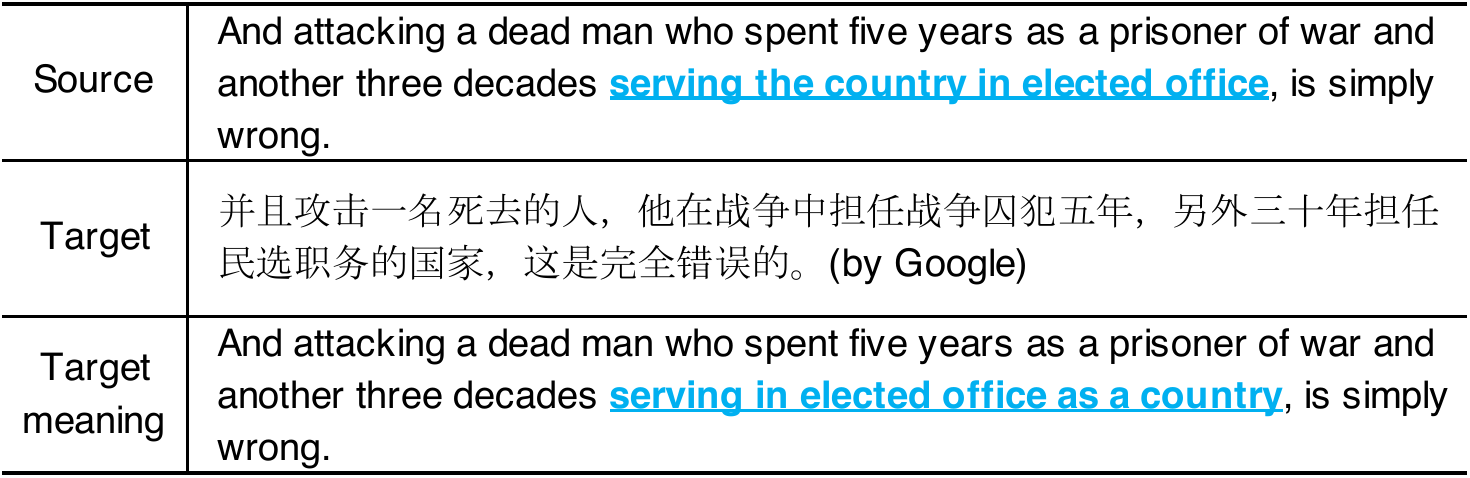}
\caption{Example of unclear logic errors detected.}
\label{fig:unclearlogic}
\end{figure}

If all the tokens/phrases are correctly translated but the sentence logic is incorrect, it is an unclear logic error. In Fig.~\ref{fig:unclearlogic}, Google Translate correctly translates "serving in the elected office" and "country." However, Google Translate generates "serving in the elected office as a country" instead of "serving the country in elected office" because Google Translate does not understand the logical relation between them. Unclear logic errors exist widely in translations given by NMT models, which is to some extent a sign of whether a model truly understands certain semantic meanings.

\subsubsection{Sentences with Multiple Translation Errors}
\begin{figure}[h]
\centering{} 
\includegraphics[scale=0.56]{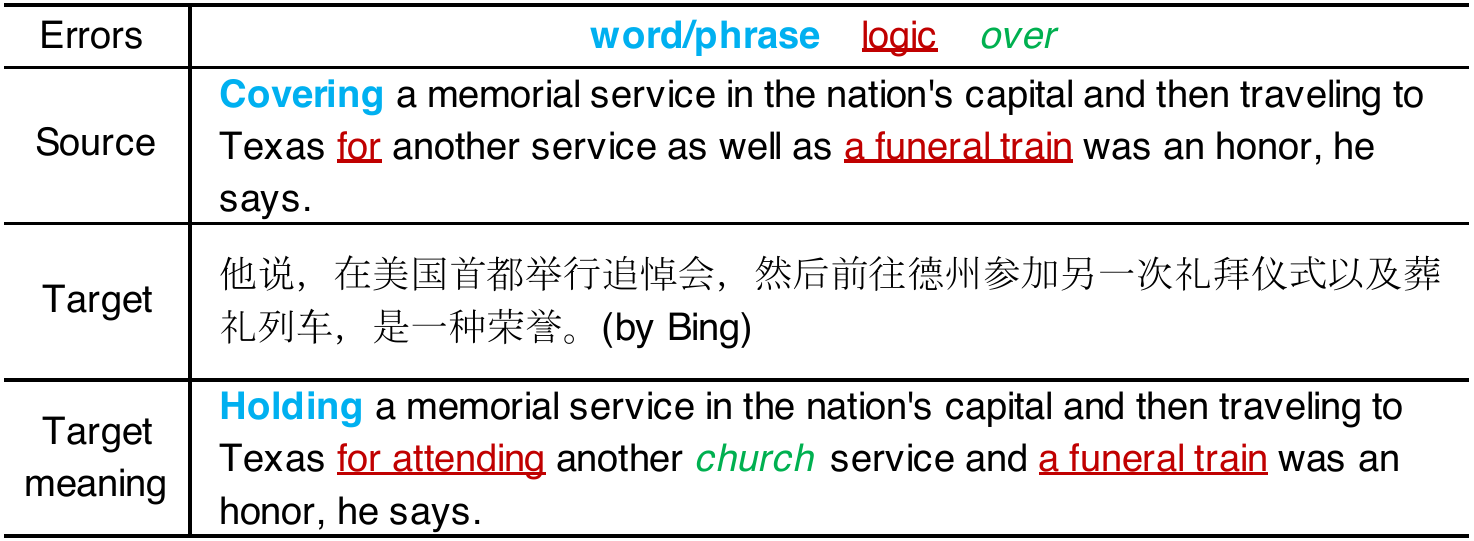}
\caption{Example of sentence with multiple translation errors detected.}
\label{fig:multiplebugs}
\end{figure}

A certain percentage of reported sentence pairs contain multiple translation errors. Fig.~\ref{fig:multiplebugs} presents a sentence pair that contains three kinds of errors. Specifically, "covering" means "reporting news" in the source sentence. However, it is translated to "holding," leading to a word/phrase mistranslation error. Additionally, "church" in the target sentence is not the translation of any words from the source sentence, so it is an over-translation error. Bing Microsoft translator also wrongly thinks the subject is "attending a funeral train." But the source sentence actually means the subject is "covering a funeral train," so it is an unclear logic error. 

\subsection{The Running Time of SIT}

\begin{figure}[]
\centering{}
\begin{tabular}{c c}

 \includegraphics[scale=0.22]{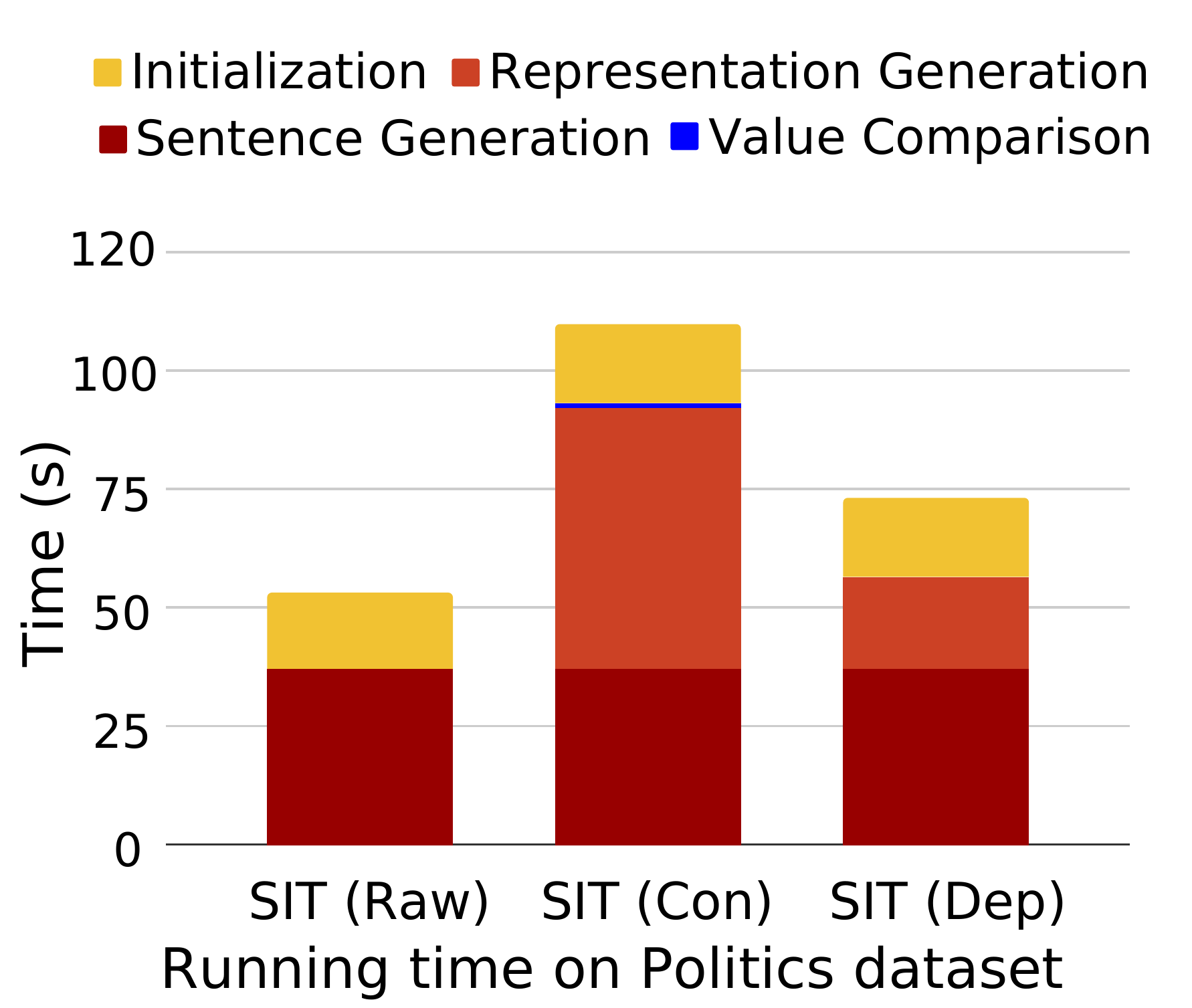} &

 \includegraphics[scale=0.22]{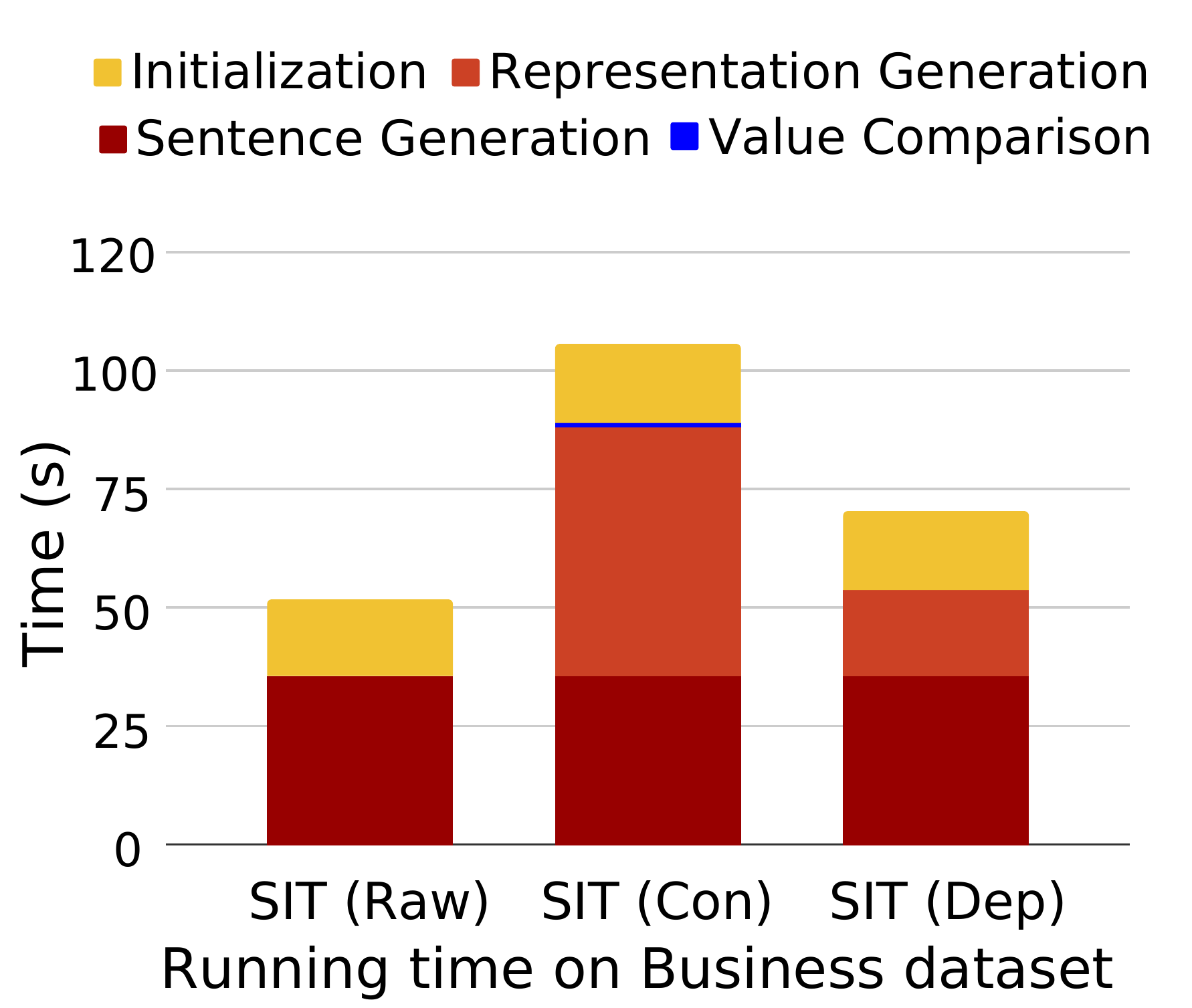} \\
\end{tabular}

\caption{Running time details of SIT (excluding translation time) in testing Google Translate.}
\label{fig:timegoogle}
\end{figure}

In this section, we evaluate the running time of SIT on the two datasets. We apply SIT with 3 different sentence structure representations to test Google Translate and Bing Microsoft Translator. We run each experiment setting 10 times and report their average as the results.
The overall running time of SIT is illustrated in Table~\ref{tab:efficiency}, and the running time of each step of SIT on Google Translate is presented in Fig.~\ref{fig:timegoogle} (Bing's result is similar). We can observe that SIT using raw sentences as structure representation is the fastest. This is because SIT (Raw) does not require any structure representation generation time. SIT using a dependency parser achieves comparable running time to SIT (Raw). In particular, SIT (Dep) uses 19 seconds to parse 2000+ sentences (as opposed to 0 seconds by SIT (Raw)), which we think is efficient and reasonable. 

\begin{table}[h]
\centering{}
\caption{Average running time of SIT on Politics and Business datasets.}
\includegraphics[scale=0.47]{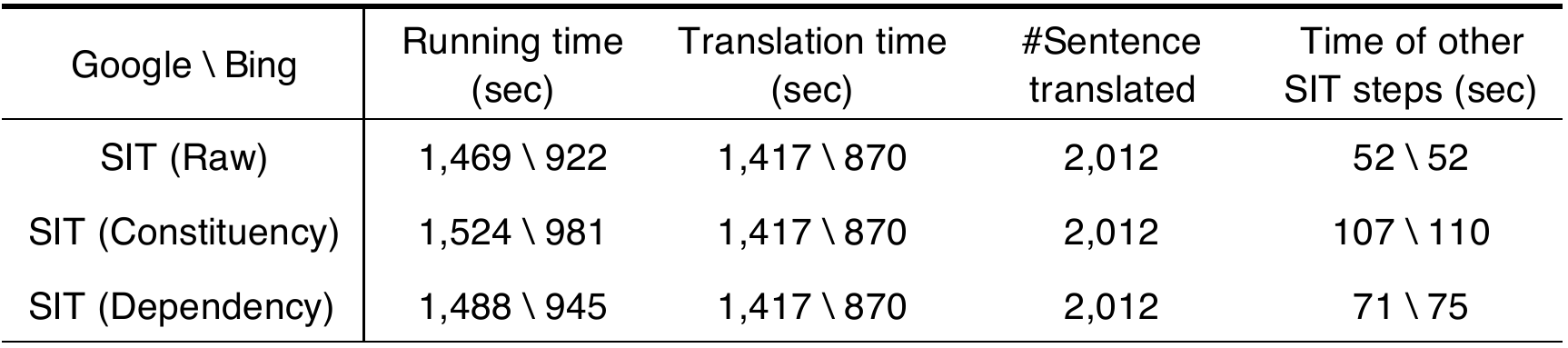}
\label{tab:efficiency}
\end{table}

In these experiments, we ran the translation step once per translation system and reused the translation results in all experiment settings since the other settings had no impact on translation time. Thus, in Table~\ref{tab:efficiency}, the \textit{Translation time} values are the same for different SIT implementations. We can observe that SIT spends most of the time collecting translation results. In this step, for each sentence, we invoked the APIs provided by Google and Bing to collect the translated sentence. In practice, if users want to test their own machine translation software with SIT, the running time of this step will be much less. As indicated in a recent study~\cite{Zhang18ACL}, current NMT model can translate around 20 sentences per second using a single NVIDIA GeForce GTX 1080 GPU. With more powerful computing resource (e.g, TPU~\cite{Wu16Arxiv}), modern NMT models can achieve the speed of hundreds of sentences translation per second, which would be about 2 magnitudes faster than in our experiments. 

The other steps of SIT are quite efficient, as indicated in Table~\ref{tab:efficiency} and Fig.~\ref{fig:timegoogle}. Both SIT (Raw) and SIT (Dep) took around 1 min and SIT (Con) took around 2 mins. Compared with SIT (Dep), SIT (Con) is slower because models for constituency parsing are slower than those for dependency parsing. We conclude that as a tool working in an offline manner, SIT is efficient in practice for testing machine translation software.


\subsection{The Impact of Distance Threshold}\label{sec:threshold}

\begin{figure}
\centering{}
\begin{tabular}{c c}
\hspace{-2ex}
 \includegraphics[width=0.23\textwidth]{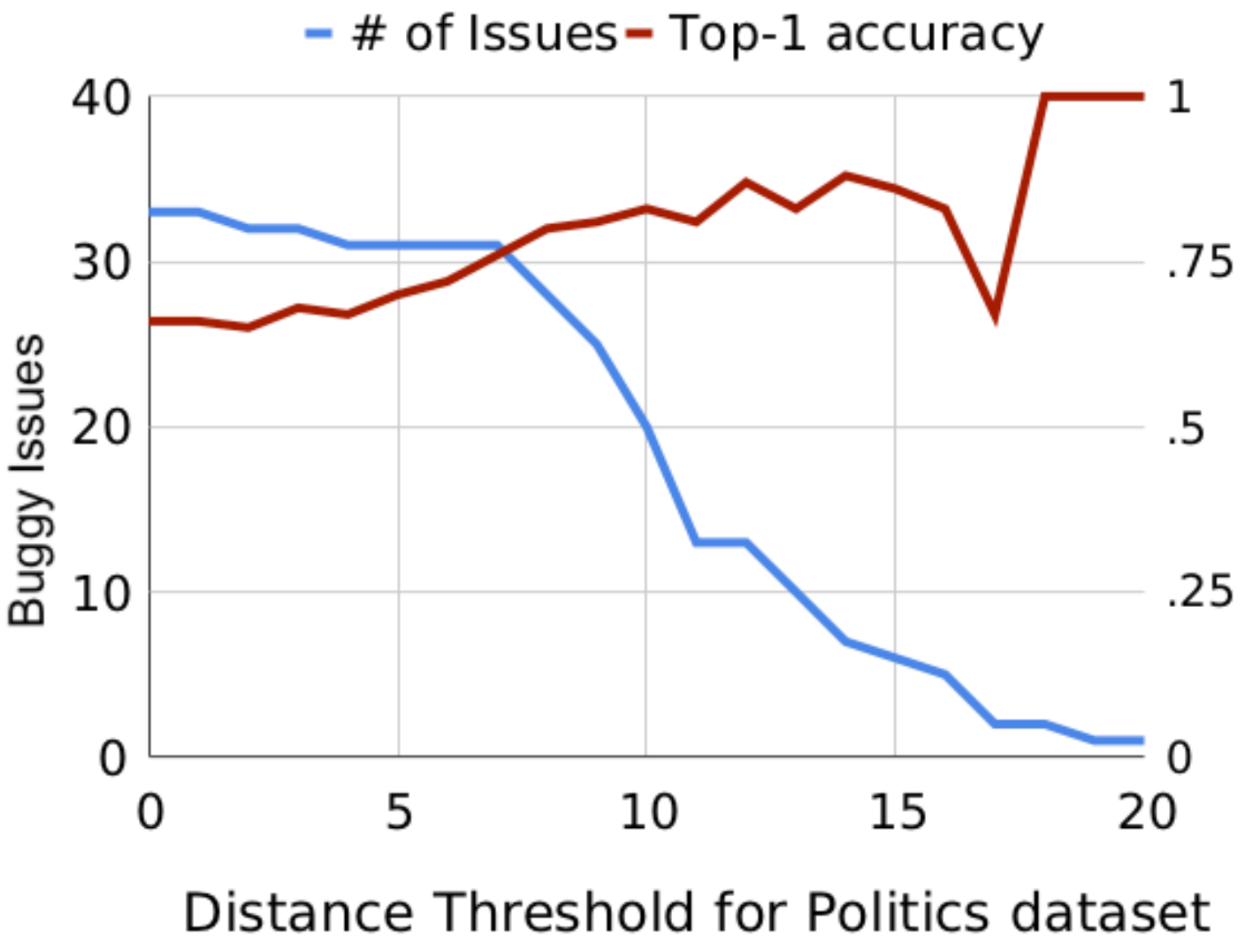} 

 \includegraphics[width=0.23\textwidth]{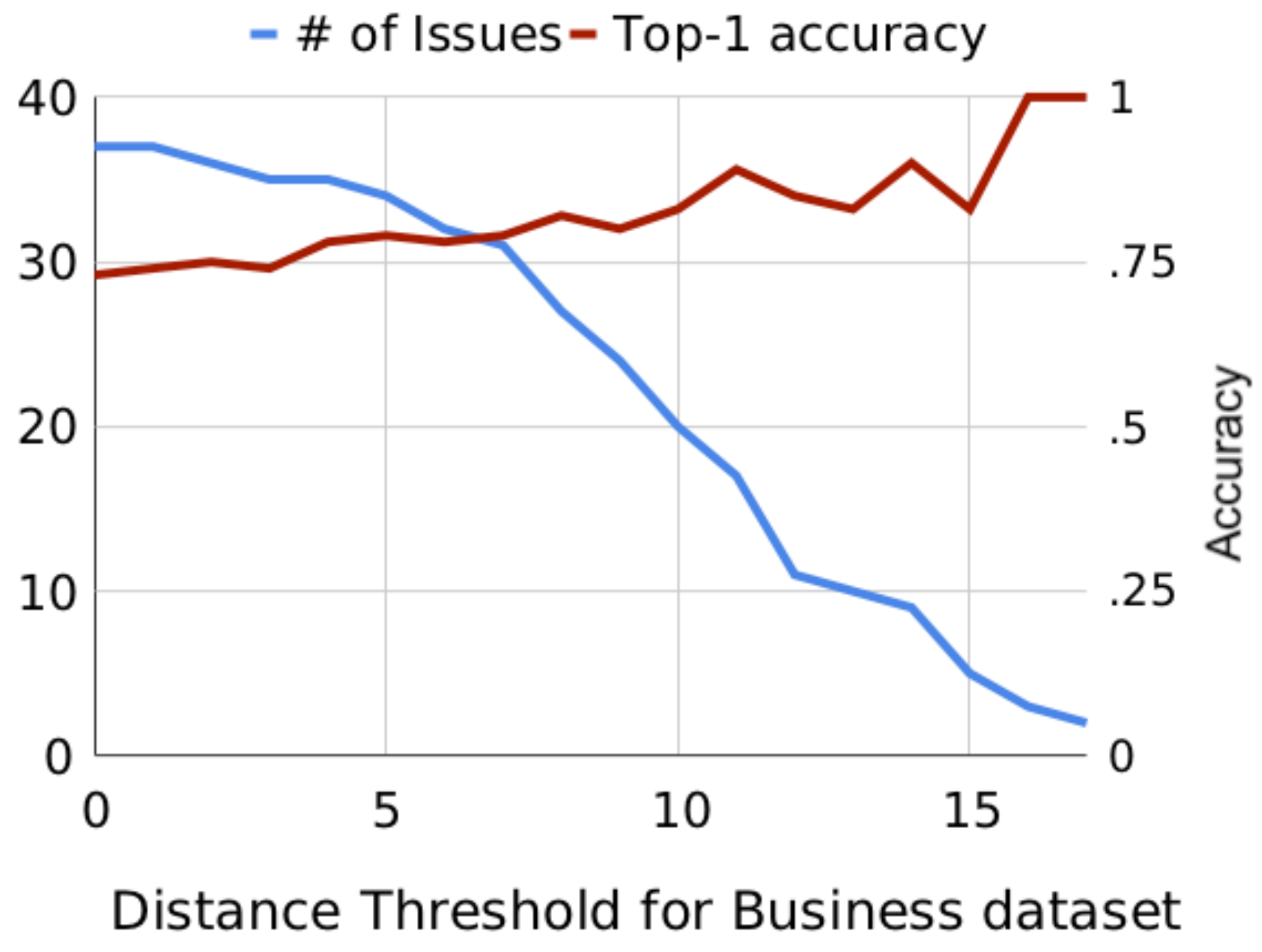} \\
\end{tabular}

\caption{Impact of distance threshold when testing Bing Microsoft Translator.}
\label{fig:parameter}
\end{figure}

SIT reports the top-k sentence pairs in an issue if the distance between the translated generated sentence and the original target sentence is larger than a distance threshold. Thus, this distance threshold controls (1) the number of buggy issues reported and (2) the top-k accuracy of SIT.
Intuitively, if we lower the threshold, more buggy issues will be reported, while the accuracy will decrease. Fig.~\ref{fig:parameter} demonstrates the impact of the distance threshold on these two factors. In this figure, SIT (Dep) was applied to test the Bing Microsoft Translator on our Politics and Business datasets with different distance thresholds. We can observe that both the number of buggy issues and top-1 accuracy remain stable when the threshold is either small or large while the values fluctuate in the middle. The impact of changing the distance threshold is similar when testing Google Translate.

Based on these results, we present some guidance on using SIT in practice. First, if we intend to uncover as many translation errors as possible, we should use a small distance threshold. A small threshold (e.g., 4 for dependency sets) works well on all our experiment settings. In particular, with a small threshold, SIT reports the most issues with decent accuracy (e.g., 70\% top-1 accuracy). We adopt this strategy in our accuracy experiments in Section~\ref{sec:accuracy}. Developers could use SIT with small distance threshold when they want to intensively test software before a release. Second, if we intend to make SIT as accurate as possible, we could use a large threshold (e.g., 15). With a large threshold, SIT reports fewer issues with very high accuracy (e.g., 86\% top-1 accuracy). Given that the number of source sentences are unlimited on the Web, we could keep running SIT with a large distance threshold and periodically report issues. Thus, we think SIT is effective and easy to use in practice.

\subsection{Fine-tuning with Errors Reported by SIT}\label{sec:finetune}
In this section, we study whether the reported buggy sentences can act as a fine-tuning set to improve the robustness of NMT models. Fine-tuning is a common practice in NMT, where training data and target data can often occupy different domains~\cite{Sennrich15ACL, Chu17ACL}. Specifically, we train an encoder-decoder model with global attention \cite{LuongPM15}---a standard architecture for NMT models---on a subset of the CWMT corpus with 2M bilingual sentence pairs~\cite{cwmt}. The encoder and decoder are unidirectional single-layer LSTMs. We train the model using the Adam optimizer \cite{Adam}, calculating the BLEU \cite{Papineni02BLEU}  score on a held out validation set after each epoch. We use the model with parameters from the epoch with the best validation BLEU score. Note that we did not use Google or Bing's translation models here because they are not open-source; however, the encoder-decoder model with attention is a very representative NMT model.

To test the NMT model, SIT is run on 40 English sentences, which are selected from the validation set of WMT'17~\cite{wmt} by removing long sentences (i.e., longer than 12 words) and ensuring that all words are in the NMT model's vocabulary. Note that since the model was not trained or validated on data from this domain, we simulated the practical scenario where real-world inputs differ from model training data. Based on these inputs, SIT successfully finds 105 buggy sentences. We label them with correct translations and fine-tune the NMT model on these 105 sentences for 15 epochs with a decreasing learning rate. After this fine-tuning, all the 105 sentences can be correctly translated. Meanwhile, the BLEU score on the original validation set used during training increases by 0.13, which, to some degree, shows that the translation of other sentences has also been improved. This demonstrates the ability to fix errors reported by SIT in an efficient and easy manner. SIT's utility on building robust machine translation software will be further elaborated in Section~\ref{sec:bugfix}.

\section{Discussions}\label{sec:discuss}

\subsection{False Positives}
While SIT can accurately detect translation errors, its precision can be further improved. In particular, the false positives of SIT come from three main sources. First, the generated sentences may have strange semantic meanings, leading to changes in the target sentence structure. For example, based on the phrase "on the way," the current implementation of SIT could generate the sentence "on the fact," which naturally has a very different translation in Chinese. Using BERT, which at the time of our experiments provided the state-of-the-art masked language model, helped alleviate this issue. Second, although the existing syntax parsers are highly accurate, they may produce wrong constituency or dependency structures, which can lead to erroneous reported errors. Third, a source sentence could have multiple correct translations of different sentence structures. For example, target sentence "10 years from now" and "after 10 years" can be used interchangeably while their sentence structures are different. To lower the impact of these factors, SIT returns the top-k suspicious sentence pairs ranked by distance to the original target sentence.


\subsection{Building Robust Translation Software}\label{sec:bugfix}
The ultimate goal of testing machine translation, similar to testing traditional software, is to build robust software. Toward this end, SIT's utility is as follows. First, the reported mistranslations typically act as early alarms, and thus developers can hard-code translation mappings in advance, which is the quickest bug fixing solution adopted in industry. Second, the reported sentences could be used as a fine-tuning set, which has been discussed in Section~\ref{sec:finetune}. Third, developers may find the reported buggy sentence pairs useful for further analysis/debugging since the sentences in each pair only differ by one word. This resembles debugging traditional software via input minimization/localization. Additionally, the structural invariance concept could be utilized as inductive bias to design robust NMT models, similar to how Shen \emph{et al.}~\cite{Shen19ICLR} introduce bias to standard LSTMs. Compared with traditional software, the debugging and bug fixing process of machine translation software is more difficult because the logic of an NMT model mainly lies in its model structure and parameters. While this is not the main focus of our work, we believe it is an important research direction for future work.


\section{Related Work}\label{sec:related}


\subsection{Robustness of AI Software}
The success of deep learning models has led to the wide adoption of artificial intelligence (AI) software in our daily lives. Despite their high accuracies, deep learning models can generate inferior results, some of which have even lead to fatal accidents~\cite{accident1, accident2, accident3}. Recently, researchers have designed a variety of approaches to attack deep learning (DL) systems~\cite{Goodfellow15ICLR, Carlini17SP, Athalye18ICML, Yang19CVPR, Xiong19CVPR,Carlini16Security, Du19Arxiv}. To protect DL systems against these attacks, excellent research has been conducted to test DL systems~\cite{Pei17SOSP,Tian18ICSE,Zhang18ASE,Ma18ASE,Ma18ISSRE, Kim19ICSE,Pham19ICSE, Xie19ISSTA, Gambi19ISSTA, Du19FSE, JZhang19Arxiv, Henriksson19AITest}, assist the debugging process~\cite{Ma18FSE}, detect adversarial examples online~\cite{Xu18NDSS, Tao18NeurIPS, Ma19NDSS, Wang19ICSE}, or train networks in a robust way~\cite{Papernot16SP, Madry18ICLR, Kannan19Arxiv, Lin19ICLR}. Compared with these approaches, our paper focuses on machine translation systems, which these works do not explore. In addition, most of these approaches require knowledge of gradients or activation values in the neural network under test (white-box), while our approach does not require any internal details of the model (black-box).


\vspace{-5pt}
\subsection{Robustness of NLP Algorithms}
Deep neural networks have boosted the performance of many NLP tasks , such as reading comprehension~\cite{Chen16ACL, Chen18Thesis}, code analysis~\cite{Iyer16ACL, Pradel18OOPSLA, Alon19POPL}, and machine translation~\cite{Wu16Arxiv, Vaswani17NeurIPS, Hassan18Arxiv}. However, in recent years, inspired by the work on adversarial examples in the computer vision field, researchers successfully found bugs produced by the neural networks used for various NLP systems~\cite{Caswell15TR, Miyato17ICLR, Iyyer18NAACL, Alzantot18EMNLP,Iyyer18NAACL,Li19NDSS,Jia17EMNLP,Mudrakarta18ACL,Ribeiro18ACL}. Compared with our approach, these works focus on simpler tasks such as text classification.

Zheng \emph{et al.}~\cite{Zheng18Arxiv} introduced two algorithms to detect two specific translation errors: under-translation and over-translation, respectively. Comparatively, our proposed approach is more systematic and not limited to specific errors. Based on the experimental results, we can find the following errors: under-translation, over-translation, incorrect modification, ambiguity of polysemy, and unclear logic. Zhou and Sun~\cite{Zhou18ASWEC} proposed a metamorphic testing approach (i.e., MT4MT) for machine translation; they followed a concept similar to structural invariance. However, MT4MT can only be used with simple sentences in a \textit{subject-verb-object} pattern (e.g., "Tom likes Nike"). In particular, they change a person name or a brand name in a sentence and check whether the translation differs by more than one token. Thus, MT4MT cannot report errors from most real-world sentences, such as the data set used in our paper. In addition, MT4MT does not propose general techniques to realize their idea. Our work introduces an effective realization via nontrivial techniques (e.g., adapting BERT for word substitution and leveraging language parsers for generating sentence structures), and conducts an extensive evaluation.

\subsection{Machine Translation}

The past few years have witnessed rapid growth for neural machine translation (NMT) architectures~\cite{Wu16Arxiv,Hassan18Arxiv}. Typically, an NMT model uses an encoder-decoder framework with attention~\cite{Zhang18ACL}. Under this framework, researchers have designed various advanced neural network architectures, ranging from recurrent neural networks (RNN)~\cite{Sutskever14NeurIPS, Luong15EMNLP}, convolutional neural networks (CNN)~\cite{Gehring17ICML, Gehring17ACL}, to full attention networks without recurrence or convolution~\cite{Vaswani17NeurIPS}. These existing papers aim at improving the capability of NMT models. Different from them, this paper focuses on the robustness of NMT models. We believe robustness is as important as accuracy for machine translation in practice. Thus, our proposed approach can complement existing machine translation research. 


\subsection{Metamorphic Testing}

Metamorphic testing is a way of generating test cases based on existing ones~\cite{Chen98Metamorphic, Segura16TSE, Chen18CSUR}. The key idea is to detect violations of domain-specific metamorphic relations across outputs from multiple runs of the program with different inputs. Metamorphic testing has been applied for testing various traditional software, such as compilers~\cite{Le14PLDI, Lidbury15PLDI}, scientific libraries~\cite{Zhang14ASE}, and database systems~\cite{Lindvall15ICSE}. Due to its effectiveness on testing "non-testable" programs, researchers have also used it to test AI software, such as statistical classifiers~\cite{Xie11JSS, Murphy08SEKE}, search engines~\cite{Zhou16TSE}, and autonomous cars~\cite{Tian18ICSE, Zhang18ASE}. In this paper, we introduce structure-invariant testing, a novel, widely applicable metamorphic testing approach, for machine translation software.

\section{Conclusion}\label{sec:con}

We have presented structure-invariant testing (SIT), a new, effective approach for testing machine translation software. The distinct benefits of SIT are its simplicity and generality, and thus wide applicability. SIT has been applied to test Google Translate and Bing Microsoft Translators, and successfully found 64 and 70 buggy issues with 69.5\% and 70\% top-1 accuracy, respectively. Moreover, as a general methodology, SIT can uncover diverse kinds of translation errors that cannot be found by state-of-the-art approaches.
We believe that this work is the important, first step toward systematic testing of machine translation software. For future work, we will continue refining the general approach and extend it to other AI software (e.g., figure captioning tools and face recognition systems). We will also launch an extensive effort to help continuously test and improve widely-used translation systems.

\section*{Acknowledgments}
We would like to thank the anonymous ICSE reviewers for their valuable feedback on the earlier draft of this paper. In addition, the tool implementation benefited tremendously from Stanford NLP Group's language parsers~\cite{stanfordcorenlp} and Hugging Face's BERT implementation in PyTorch~\cite{BERTpytorch}.
\balance
\bibliographystyle{ACM-Reference-Format}
\bibliography{References}


\end{document}